\documentclass[10pt]{article}
\usepackage{graphicx}% Include figure files

\usepackage{epsfig}
\usepackage{latexsym}
\usepackage{amsmath}

\begin{document}

\pdfoutput=1

 %\begin{frontmatter}

\title{DISCRETE MODEL OF IDEOLOGICAL STRUGGLE ACCOUNTING FOR MIGRATION}
\author{
NIKOLAY K. VITANOV $^a$\footnote{
$e$-$mail$ $address$:
vitanov@imbm.bas.bg }
\\
MARCEL AUSLOOS $^b$\footnote{
$e$-$mail$ $address$:
marcel.ausloos@ulg.ac.be } 
\\
GIULIA ROTUNDO $^{c}$\footnote{  $e$-$mail$ $address$: giulia.rotundo@uniroma1.it}
\\ $^{a}$ Institute of Mechanics, Bulgarian Academy of Sciences, 
\\Acad. G. Bonchev Str., Bl. 4, \\Sofia, 1113, Bulgaria 
\\ $^{b}$ %{\it now at}
 GRAPES, Beauvallon Res., r. Belle Jardiniere, 483\\
Liege Angleur, B-4031, Euroland
\\ $^c$ Faculty of Economics, University of Tuscia, via del Paradiso 47\\
Viterbo, I -01100, Italy  }

\date{\today}
\maketitle
\vskip 0.5truecm

\begin{abstract}
A discrete in time   model of ideological competition is formulated taking into account  population
migration.  The model is based on interactions between  $global $ populations 
of non-believers and followers of different ideologies. The complex dynamics of 
the attracting manifolds is investigated.  
 Conversion from one ideology to another by means of  (i) mass media influence and (ii) 
interpersonal relations is considered.  Moreover a different birth rate is assumed for different ideologies, 
the rate being assumed to be positive for the reference population, made of initially non-believers.  
Ideological competition can happen in one or several regions in space. In the latter case, migration of 
non-believers and adepts is allowed;  this leads to an enrichment of the ideological dynamics. Finally,
the current ideological situation in the Arab countries and China is commented upon 
from the point of view of the presently developed mathematical model. The massive forced conversion by 
Ottoman Turks in the Balkans is briefly discussed.
\end{abstract}

%\keywords{social dynamics; ideological struggle; migration.}

\section{Nonlinear systems, population dynamics, and opinion formation}
Competition among ideologies for the people minds is an  interesting phenomenon 
that can lead to large consequences for the social structure and for the 
social evolution of populations.
Religious wars in the past or   social revolutions during the XX-th century are
closely connected to  such  ideological competitions.  Ideological competition is
possible because   people might and can accept or drop ideas and systems of ideas
and thereafter act according to the newly adopted system of ideas \cite{vdm10,hinich}. 
Population of believers or non-believers do form. The number of individuals in 
these populations can change with  time. Thus the ideological competition is closely 
connected to population dynamics state but the human ideological struggle has its 
specific features:  animals cannot change  their kind (zebras cannot become lions; 
electrons cannot become bosons) but  people  can  (though sometimes may not) 
change their ideology.
\par
The systems studied as  population dynamics are among the best examples of
nonlinear systems in quantitative social and natural sciences \cite{MontrollBadger74}. 
Different kinds of nonlinearities arise in the model equations 
of such systems, for  example  (i) because of interactions among the individuals 
or the populations or (ii) because of  limitations  in the environment 
\cite{brauer,vjd09,hastings,vjd09a,okubo,vdk06}.  Such nonlinear model systems  
require applying  methods of nonlinear dynamics \cite{kaplan,v98,mees,vj06,psv07}, 
chaos theory \cite{kiel,dv01,bv02,dv04}, 
theory of stochastic processes \cite{gard,vank,kv04}  and
even variational and other methods of the theory of turbulence \cite{v00,v00a,v00b}. 
\par
In this paper,   ideological struggle will be considered to occur on a 
two-dimensional surface (regions in a country or regions containing an arbitrary 
number of countries). We shall assume that the population of the corresponding
territory is divided  into sub-populations of followers of several ideologies and one
population of non-believers. Any newborn, whatever  each parent ideology 
is necessarily at first a non-believer; any ideology is acquired later.  
This assumption has as a consequence that the rate of 
change of the number of individuals by birth and death can be positive only for
the population of  non-believers. The corresponding rate for the sub-populations
of the followers of the different ideologies is necessarily  negative as,  for these, only death
is possible.
\par 
What will be apparently new in our model is the possibility of  migration of groups of
non-believers or followers of ideologies. Indeed it is going to be assumed that non-believers 
or followers of different ideologies can move from one region to another one, even 
if the regions do not have a common boundary: thus allowing motion by airplanes or 
due to a transit crossing of a region by means of train, ship, bus, car, bicycle, horse, camel, 
donkey, etc.. This movement can change its intensity in  
time; moreover the birth rates and conversion coefficients  are allowed to change 
as well. The result of this is a very complicated dynamics in  time and  space. 
\par 
 The  model system of nonlinear 
equations is taken as a system of difference equations which means that one 
assumes discrete time. One  fundamental reason for this is  our wish that the 
model  be close to the true situation as much as possible, in the sense that  data 
is usually collected for  discrete time intervals and for a finite number of 
regions. These two features of the real  social life are  thus accounted for in the
model below.
\par
The paper is organised as follows. In   Sect. \ref{singleregion}, the 
system of model equations is formulated for the case of a single region, for 
simplicity and for outlining the notations. The possibility of time change of the coefficients of the model equation 
leads us to the notion of {\it special points}. Special points of the system of  these model 
equations are calculated on the basis of the rules used  for calculating so called  fixed 
points. However as the coefficients of the system vary in  time, the points are not  
truly fixed but move in the phase space. Because of this, we use for such points 
the name special points. If the coefficients of the model system of equations do 
not change in time, the special points become  usual fixed points.   
\par
Sect. \ref{tworegions} is devoted to a discussion of the model 
for the case of two, obviously adjoining, regions. 
The possibility of migration  being allowed, the dynamics become more complicated.
Two particular cases are discussed: massive leaving of non-believers from one of 
the regions to the other, in Sect. \ref{massleaving} and
massive invasion by followers of some ideology from one of the two regions to the other  in Sect. \ref{massinvasion}. Sect. \ref{fullmodel} is devoted to the 
formulation of the model for an arbitrary number of regions and an arbitrary number 
of ideologies. On the basis of the general model we discuss the possibilities for 
the strengthening or weakening
of an official ideology. Sect. \ref{remarks} contains several pertinent remarks.  
Examples are taken from recent and non-recent history : Arab countries and China, and
the spreading of Islam in Europe from the beginning of the 15th century though the 
reader can be aware of other cases according to its own knowledge of history. Other discussions and conclusions are found based on a summary in Sect. \ref{conclusions}.
\section{Single region}\label{singleregion}
\par
\subsection{Model}\label{model}
In order not to overload the basic equations with many terms, let us first  present 
the model and some discussion in the case of  a single region. In this region,  
consider that there are $n+1$ population types, or ideologies:  one specific population of
non-believers, or having $no$  opinion at all about some subject, with spatial density 
$\rho_0^{(t)}$ of members at   time $t$, and $n$ populations with spatial 
densities $\rho_i^{(t)}$ of followers of different ideologies ($i=1, ..., n)$  about some given question
(the measure of all densities is people per square kilometer). There is no restriction to considering 
only $political$  $ideologies$;  a religion  can be taken as an ideology as well.\footnote{Languages are not 
here considered to be ideologies, due to our consideration 
that one can only belong to one ideology at a given time, thus we exclude multilingual 
populations for our present considerations.}  Notice that $i$ can be considered as an index indicating
the component of the {\it degree of freedom vector} characterizing an individual. However, in view of
being more general, one can consider that some $i$ corresponds to a population, having some mere 
(so called scientific) degree of freedom for characterising part of the population, i.e. e.g. 
characterised by  by a political $and$ a philosophical attitude, like distinguishing between ($i=1$) 
socialist Buddhists and ($i=2$) liberal Buddhists, or ($i=3)$  socialists jews and ($i=4)$ liberal jews, etc.\footnote{
Of course other notations can be used, like $\rho_{i,j,k,l,\dots}$ where $i$ indicates the ideology, $j$ the religion,
$k$ the sex, $l$ the social status, $\dots$, etc.} 
Thus the total density at   time $t$ is 
\begin{equation}\label{a1}
\rho^{(t)} = \rho_0^{(t)} + \sum_{i=1}^n \rho_i^{(t)}
\end{equation}
\par
The model (difference) equations for describing the interactions among the above  
$n+1$ populations are assumed to be of the Lotka-Volterra type
\begin{eqnarray}\label{a2}
\rho_i^{(t+1)} - \rho_i^{(t)}=r_i^{(t)} \rho_i^{(t)} + \sum_{j=0}^n f_{ij}^{(t)} \rho_j^{(t)} +
\sum_{j=0}^n b_{ij}^{(t)} \rho_i^{(t)} \rho_j^{(t)}
\end{eqnarray}
where $r_i^{(t)}$ is the  ''biological''  rate for changing the density of the 
corresponding population by births and deaths. In other words $r_i^{(t)}$  
accounts for the finite duration of the life of the people.
In the present work, no time delay, memory  or anticipation is considered between 
various causes and effects. The change in population number occurs at  each unit of  time $t$  which is discrete. 
As already mentioned in the previous section below we assume 
that only the population of non-believers  can have a positive 
 $r_0^{(t)}$ rate: thus  we assume that   (i) all newborn babies are non-believers and  (ii) for this population, the
number of births can be larger than the number of deaths.  For the other populations
of believers, the rates $r_i^{(t)}$, ($n\ge i \ge 1$)  are taken to be negative, since only death is 
possible there. % (and of course conversion as we shall see just below). 
\par
In addition to births and deaths, we assume in the second and third term of the equation r.h.s. that $two$ kinds of conversion exist in the 
system of populations. The first kind of conversion is  triggered by the mass media:
newspapers, Internet, TV, radio, etc. In the model equation for the $i$-th population, 
the mass media conversion is accounted for by the 
term $f_{ij}^{(t)} \rho_j^{(t)}$ where $f_{ij}^{(t)}$ is the corresponding coefficient 
of conversion influencing the number of   followers of the $j$-th ideology  who are 
converted to the $i$-th ideology because of mass media  influence.
\par
The second kind of conversion is the contact conversion that happens, e.g.,  after 
conversations between members of different ideologies. 
We assume for simplicity that the result from this 
kind of conversion is proportional to
the densities  of the  conflictual  contacting ideologies. Thus
this kind of conversion is modelled by the term $b_{ij}^{(t)} \rho_i^{(t)} \rho_j^{(t)}$, 
where
$b_{ij}^{(t)}$ is a coefficient that accounts for the intensity of the conversion.

\par
The dynamics of the system of $n+1$ populations as described by Eq. (\ref{a2}) 
can be discussed in an $n+1$-dimensional Cartesian phase space. The coordinates on  the 
axes  spanning this space correspond to the densities  $\rho_0, \rho_1, \dots, \rho_n$ of the 
adepts of the corresponding ideology. The coordinates  $\rho_0, \rho_1, \dots, \rho_n$   
define a point in this phase space denoting the  social state of the investigated system
at a given   time.  As the time $t$ evolves,  the system state follows a trajectory: the  so 
called {\it system trajectory}. This system trajectory can be attracted to different 
kinds of manifolds in the phase space. These manifolds are called attractors: they can 
be  points, or more complicated structures.  Most of the time the system is on its way to  
some attracting manifold, which in most cases  can be some equilibrium state.
\par
\subsection{Spreading of ideologies}\label{spreading}
Let us now discuss the simplest case, i.e. spreading of one ideology in one area  due 
to opinion competition.  For simplifying the presentation, let only two 
populations/opinions/ideologies existing:  (i) the non-believers $\rho_0$ 
and (ii) the followers  $\rho_1$   
of ideology $1$. The model equations read
\begin{eqnarray}\label{a3}
\rho_0^{(t+1)} - \rho_0^{(t)}=r_0^{(t)} \rho_i^{(t)} + \sum_{j=0}^1 f_{0j}^{(t)} \rho_j^{(t)} +
\sum_{j=0}^1 b_{0j}^{(t)} \rho_0^{(t)} \rho_j^{(t)}
\end{eqnarray}
\begin{eqnarray}\label{a4}
\rho_1^{(t+1)} - \rho_1^{(t)}=r_1^{(t)} \rho_1^{(t)} + \sum_{j=0}^1 f_{1j}^{(t)} \rho_j^{(t)} +
\sum_{j=0}^1 b_{1j}^{(t)} \rho_1^{(t)} \rho_j^{(t)}
\end{eqnarray}
Without any loss of generality,  let also $f_{00}=f_{11}=0$ and $b_{00}=b_{11}=0$,   i.e.  the  mass media influence  and binary conversion do  not increase the number of members
of both populations  {\it per se} through some miraculous multiplication/addition. Then the system of equations  reduces to
\begin{equation}\label{a5}
\rho_0^{(t+1)} - \rho_0^{(t)}=r_0^{(t)} \rho_0^{(t)} + f_{01}^{(t)} \rho_1^{(t)} + b_{01}^{(t)} 
\rho_0^{(t)} \rho_1^{(t)}
\end{equation}
\begin{eqnarray}\label{a6}
\rho_1^{(t+1)} - \rho_1^{(t)}=r_1^{(t)} \rho_1^{(t)} + f_{10}^{(t)} \rho_0^{(t)} + b_{10}^{(t)} 
\rho_1^{(t)} \rho_0^{(t)}.
\end{eqnarray}
Remember that $r_1^{(t)}$ is imposed to be negative and $r_0^{(t)}$   positive.
\par
The special phase space points connected to the dynamics of the
system, Eqs. (\ref{a5}) and (\ref{a6}),  are obtained 
when the left-hand sides of the  two equations (\ref{a5}), (\ref{a6}) are equal to $0$.
We shall call such points special points instead of fixed points because of the following
reason. Fixed points do not move in the phase space, but the special points can move in
this space, because   the system parameters can change with  time.  Only the special point corresponding to the extinction of all
populations is a fixed  point for all values of the system parameters. 
\par
The  (two) special points for the system of equations  (\ref{a5}) and (\ref{a6}) are
\begin{equation}\label{a7a}
\rho_{0}^{(t)}=\rho_{1}^{(t)}=0  
\end{equation}
and the non-trivial special point  \begin{equation}\label{a7b}
\rho_{0}^{(t)}=\frac{f_{10}^{(t)} f_{01}^{(t)} - r_1^{(t)} r_0^{(t)}}{r_0^{(t)}
b_{10}^{(t)} - b_{01}^{(t)} f_{10}^{(t)}}, \;\;\;\; 
\rho_{1}^{(t)} = \frac{f_{10}^{(t)} f_{01}^{(t)} - r_1^{(t)} r_0^{(t)}}{r_1^{(t)}
b_{01}^{(t)} - b_{10}^{(t)} f_{01}^{(t)}}.
\end{equation}
N.B. If $r_0^{(t)}b_{10}^{(t)} - b_{01}^{(t)} f_{10}^{(t)}=0$ or $r_1^{(t)}
b_{01}^{(t)} - b_{10}^{(t)} f_{01}^{(t)} =0$, there exists a single special
point $\rho_{0}^{(t)}=\rho_{1}^{(t)}=0$.
\par
The first special point  $\rho_{0}^{(t)}=\rho_{1}^{(t)}=0$  corresponds to the extinction 
of the (two) populations. We note that this special point, when reached after some 
trajectory depending on the initial conditions, does not move in phase space: 
it is a fixed point. The second special point is much more interesting. In  
general,  the coefficients in the model equations (\ref{a5}), (\ref{a6}) can change their 
values in the course of  time. Thus, the second special point can move in  the phase space.
As the phase point corresponding to the system state moves,  one can observe  an 
interesting $hunt$ for equilibrium: the phase point 'hunts' the special point.
The dynamics of the system trajectory in the phase 
space can become quite complicated, as seen in Fig. 1,  for a particular numerical example.
Changing $r_1$ leads to additional movement of the special point as shown in Fig. 2 . 
\begin{figure}[th]
\vskip1cm
%\begin{center}
\centerline{\psfig{file=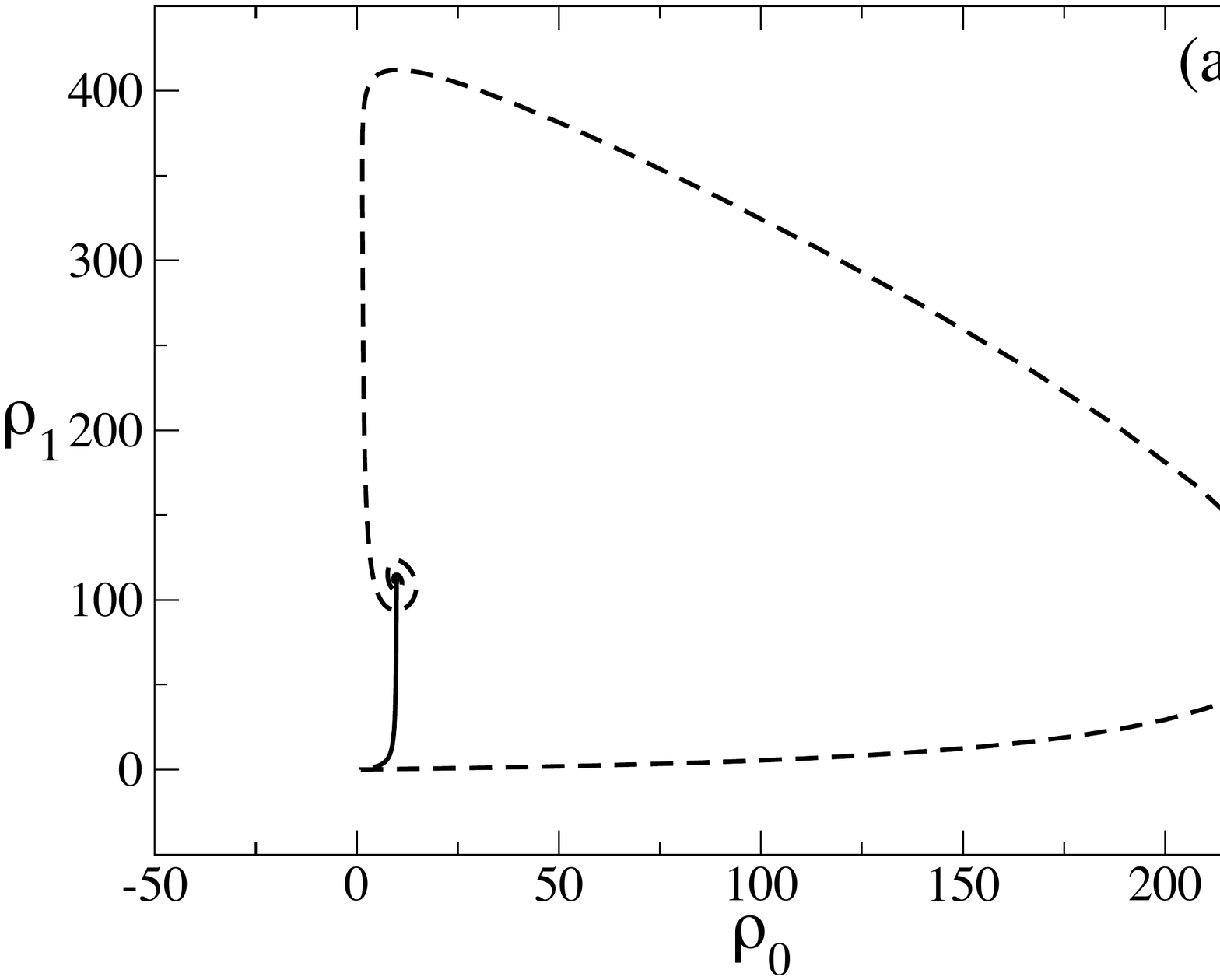,width=6.5cm}}
%\end{center}
\vspace*{8pt}
\caption{   Illustration of the system trajectory and the special points
in the phase space for the case of a single area and two populations, - non-believers and
followers of an ideology. Solid line: movement of the non-zero special point given by
Eq.(\ref{a7b}). Dashed line: system trajectory calculated on the
basis of Eq.(\ref{a5}) and Eq.(\ref{a6}). The parameters are appropriately chosen 
in order to illustrate how the system phase point 'hunts' and 'catches' the  (non trivial) special
point ($\rho_0^{(t)}, \rho_1^{(t)}$) as $t \rightarrow \infty$.
Initial values: $\rho_0 =0.98$, $\rho_1=0.02$. The constant parameters are:
$f_{10}^{(t)}=0.0002$; $f_{01}^{(t)}=0.0001$; $r_1^{(t)} = -0.001$; $b_{01}^{(t)}=-0.0001$;\;
$b_{10}^{(t)}=0.0001$.  All such parameters  are assumed to be constant. The  only time 
changing parameter is: $r_0^{(t)}=0.01 \tanh(0.01 t)$. The maximum value of $t$ used   is
$t=40 000$.}
\end{figure}

\begin{figure}[th]
\vskip1cm
\centerline{\psfig{file=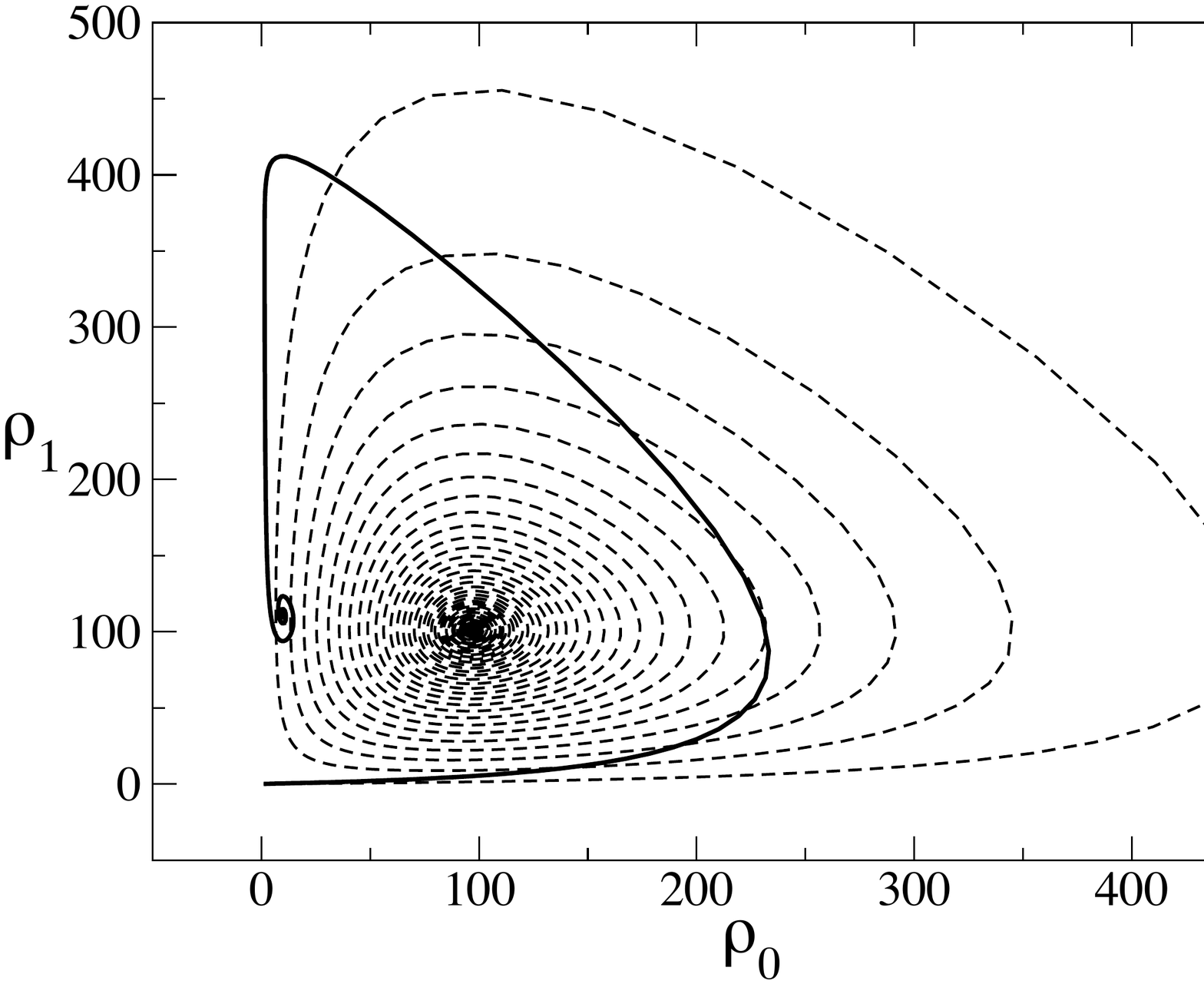,width=6.5cm}}
%\end{center}
\vspace*{8pt}
\caption{    Effect of the change of
$r_1$.  Solid line: the corresponding trajectory from Fig.1. Dashed line:
the system trajectory for  $r_1^{(t)} = -0.01[1-
0.01 \tanh(0.001t)]$.  Illustration of the system trajectory and the special points
in the phase space for the case of a single area and two populations, - non-believers and
followers of an ideology. Solid line: movement of the non-zero special point given by
Eq.(\ref{a7b}). Dashed line: system trajectory calculated on the
basis of Eq.(\ref{a5}) and Eq.(\ref{a6}). The parameters are appropriately chosen 
in order to illustrate how the system phase point 'hunts' and 'catches' the  (non trivial) special
point ($\rho_0^{(t)}, \rho_1^{(t)}$) as $t \rightarrow \infty$.
Initial values: $\rho_0 =0.98$, $\rho_1=0.02$. The constant parameters are:
$f_{10}^{(t)}=0.0002$; $f_{01}^{(t)}=0.0001$; $b_{01}^{(t)}=-0.0001$;\;
$b_{10}^{(t)}=0.0001$.  All such parameters  are assumed to be constant. The  only time 
changing parameter is: $r_0^{(t)}=0.01 \tanh(0.01 t)$. The maximum value of $t$   is
$t=40 000$.}
\end{figure}

\par
 As we can see the system trajectory (marked with the dashed line in Fig. 1) has to make 
a (long) journey through the phase space before finding and reaching the non-zero state 
described by the moving special point.  This occurs despite the fact that at the initial  time 
the system state   and the special point are not very far one from another. Thus, the 
possibility that a time change in system coefficients enriches the dynamics in comparison to 
the case of constant coefficients of the model equations is easily demonstrated. 
\par
Quite interesting are the phenomena connected to the linear stability of the two special
points from Eqs.(\ref{a7a})-(\ref{a7b}). The eigenvalues connected to the stability of the special
point $(0,0)$, which is  a fixed point,  are
\begin{equation}\label{a8}
\lambda_{1,2}^{(t)}=\frac{1}{2} \left[ r_0^{(t)} + r_1^{(t)} \pm \sqrt{(r_0^{(t)}- 
r_1^{(t)})^2+4 f_{01}^{(t)} 
f_{10}^{(t)}} \right ].
\end{equation}
Let us assume that the coefficients connected to the mass media conversion are both 
positive, i.e.  $f_{01}^{(t)}>$ and $f_{10}^{(t)}>0$. Then the two eigenvalues are real. If 
the system parameters are constant in time and
$\mid \lambda_1 \mid <1$ and $\mid \lambda_2 \mid <1$ the special point is stable.
But as the parameters can vary with   time the  condition $\mid \lambda_1 \mid <1$ and 
$\mid \lambda_2 \mid <1$ can be violated and the
point may become unstable for some  time interval. If it is unstable, it cannot be reached.

\begin{figure}[t]
\vskip1cm
%\begin{center}
%\includegraphics[scale=0.45]{1area_2pop_complex_motion.eps}\label{F2a}
%\includegraphics[scale=0.45]{movement_fixp.eps}\label{F2b}
%\end{center}
\centerline{\psfig{file=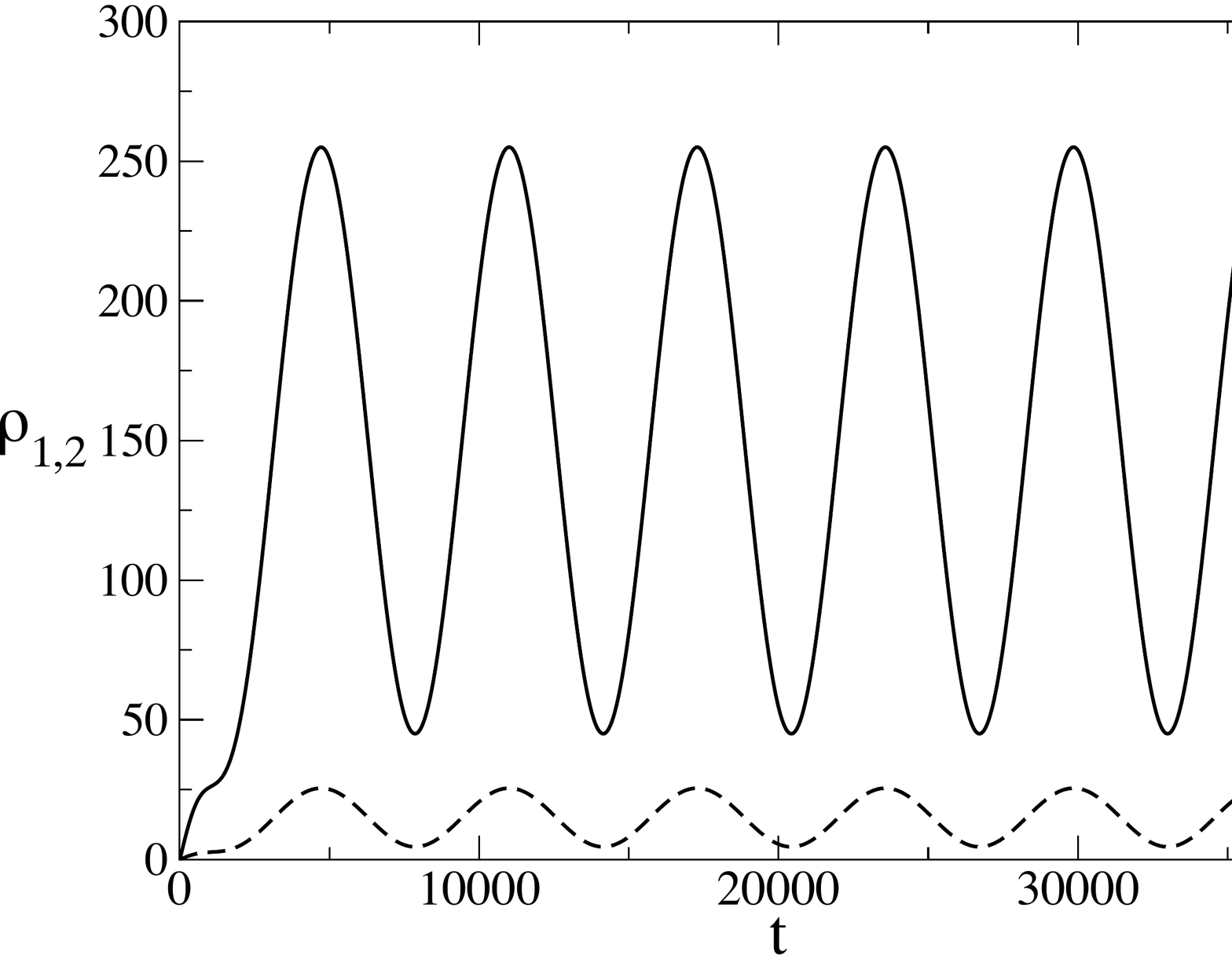,width=8.5cm}}

 \caption{ Illustration of a periodic motion of the non-trivial special point, i.e., 
 motion of the components $\rho_{0}^{(t)},\rho_{1}^{(t)}$ of the special point 
from Eq.(\ref{a7b}). Solid line: $\rho_{0}^{(t)}$; dashed line: $\rho_{1}^{(t)}$.   Initial values:    $\rho_0^{(1)}=6.0 \cdot 10^{-2}$; 
$\rho_1^{(2)}=2.0 \cdot 10^{-1}$;  the parameters  $f_{10}= 10^{-4}$; $f_{01}=0$; $r_1=-10^{-3}$ ; 
$b_{01}=-0.1$; $b_{10}=0$ are assumed to be constant, but $r_0=1.5 \tanh(10^{-3} t) - 1.05 \sin( 10^{-3} t)$. N.B. the $y$-axis should read $\rho_{0,1}$ and not $\rho_{1,2}$. }
\end{figure}

\begin{figure}[t]
\vskip1cm
%\begin{center}
%\includegraphics[scale=0.45]{1area_2pop_complex_motion.eps}\label{F2a}
%\includegraphics[scale=0.45]{movement_fixp.eps}\label{F2b}
%\end{center}
\centerline{\psfig{file=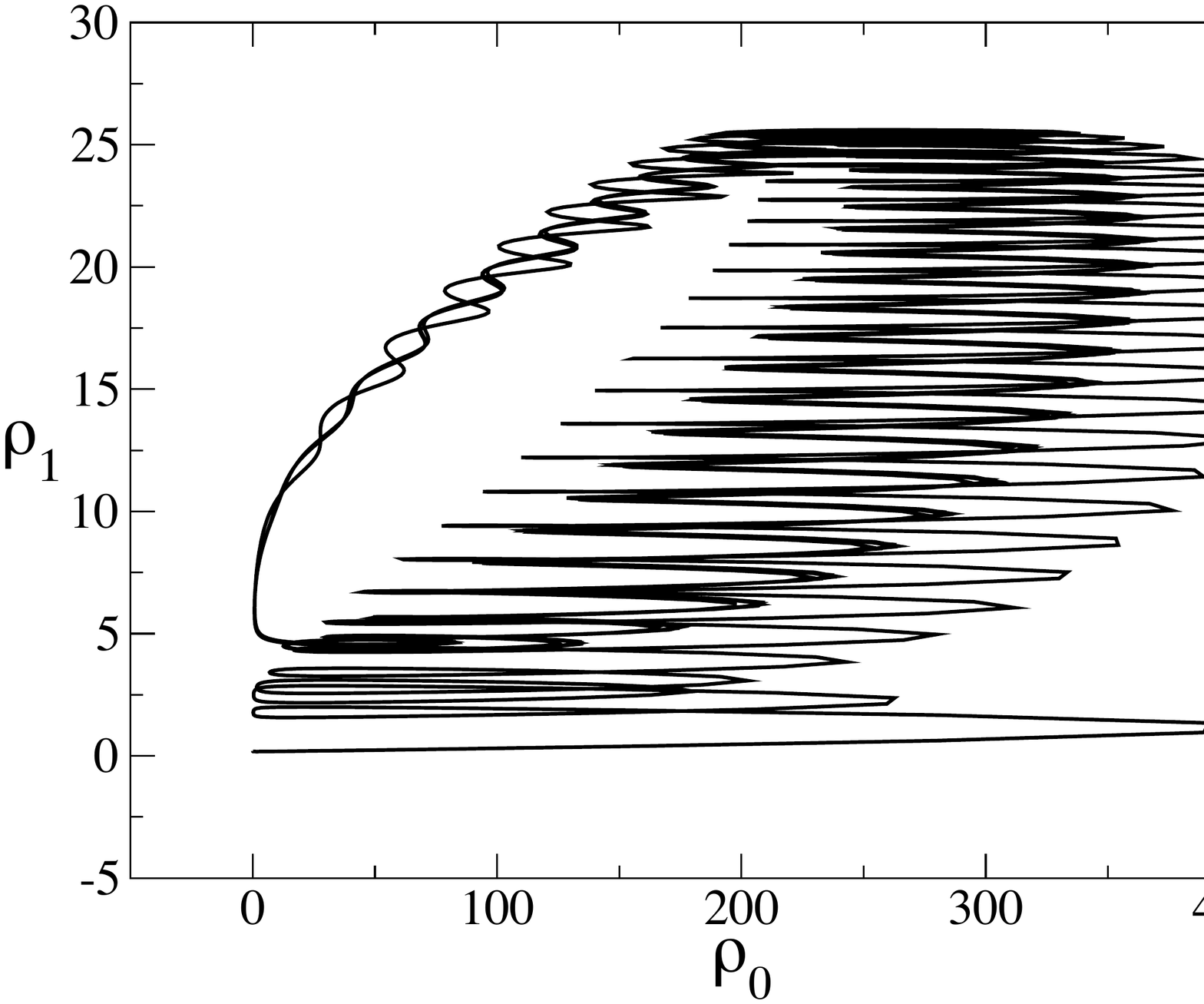,width=8.5cm}}

 \caption{ Illustration of the corresponding complicated trajectory of the system 
in the phase space  
from Eq. (\ref{a7b}). All values of the parameters and quantities are chosen 
in order to make the illustration more visible.    Initial values:    $\rho_0^{(0)}=6.0 \cdot 10^{-2}$; 
$\rho_1^{(0)}=2.0 \cdot 10^{-1}$;  the parameters  $f_{10}= 10^{-4}$; $f_{01}=0$; $r_1=-10^{-3}$; 
$b_{01}=-0.1$; $b_{10}=0$ are assumed to be constant, but $r_0=1.5 \tanh(10^{-3} t) - 1.05 \sin( 10^{-3} t)$ }
\end{figure}

\par
In addition a special point may move periodically; whence any  attempt
 for the system trajectory  to catch the moving special point 
can lead to a complex motion of the phase trajectory
in the phase plane. This can be illustrated on the basis of the following
example. Let $f_{01}=0$ and $b_{10}=0$ and 
fix  the values of parameters as mentioned in the caption of   Figs. 3-4.  Let all 
parameters be constant, except for $r_0$ which oscillates.  The result  is an oscillating special point; 
the system trajectory moves in a complex way in the phase space (see Fig. 3). 
\par
Of course, there can be more than one ideology, and the so called non-believers,  present in the  region.
Then the existing ideologies can compete for  adepts, such that the dynamics of the state point in the phase
space can become more complicated.  An illustration  of such a case is shown in 
Fig. 5    for a particular numerical example.

\begin{figure}[th]
%\vskip-3cm
%\begin{centerj}
\centerline{\psfig{file=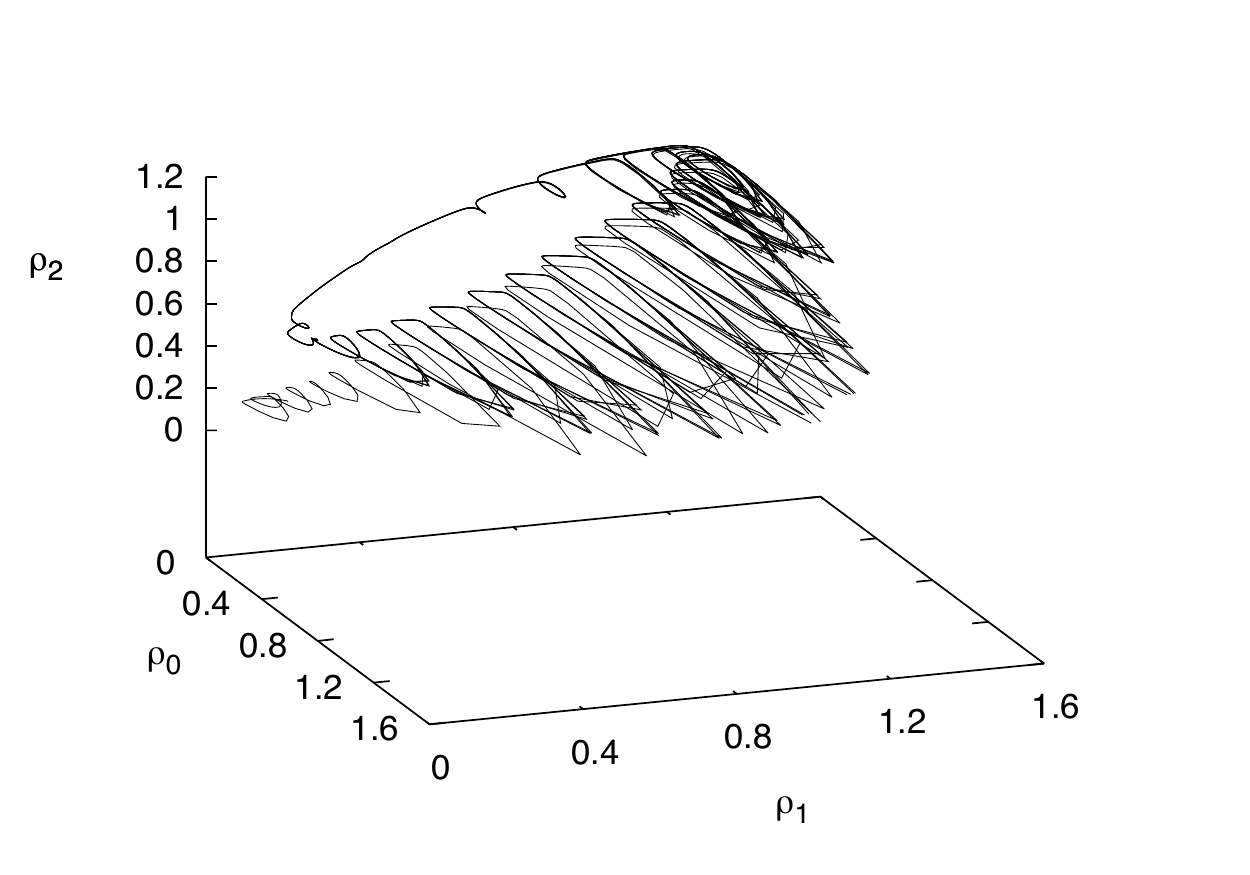,width=12cm}}
%\end{center}
\vspace*{8pt}
\caption{Illustration of the trajectory of a system made of three populations, i.e. 
non-believers and followers  of two distinct  ideologies.
The initial values for the run are:  $\rho_0=1;\rho_1=0.2; \rho_2=0.1$. Other
parameters have the constant values:
$f_{10}=10^{-2}; f_{12}=10^{-5}; f_{01}=0; f_{02}=10^{-6}; f_{20}=10^{-3};
f_{21}=10^{-3};\; ; \; r_1=-10^{-3}; r_2=-10^{-3};\; b_{01}=-0.1; b_{10}=0;
b_{02}=-10^{-4}; b_{12}=10^{-6}; b_{20}=10^{-5}; b_{21}=-10^{-4}$, but  $r_0 = 1.5
\tanh(0.001 t)-1.05 \sin (0.001 t) + 0.02 \sin (0.03 t)$. The trajectory is
calculated up to $t= 40 000$}
\end{figure}
\section{The case of two regions }  \label{tworegions}
Consider two regions, $  \mathrm{I}$ and $ \mathrm{II}$,  and some possibility for migration of believers or non-believers 
between the regions. Let us note the $n+1$ densities of populations/opinions in the
first region  as $\rho_{\mathrm{I,}0}, \rho_{\mathrm{I,}1},\ ..., \rho_{\mathrm{I,}n}$ and 
the densities of populations
in the second region  as $\rho_{ \mathrm{II},0}, \rho_{ \mathrm{II},1},  \dots, \rho_{ \mathrm{II},n}$ (some of $\rho_{ \mathrm{I},i}$ and
$\rho_{ \mathrm{II},j}$ can be $0$). Let it be conventionally accepted from now on  that the label before the comma in the index refers 
to the region, while the index after the comma refers to the ideology type or population. The leaving  from and coming to a  region are 
accounted for by introducing new terms in the
model equations  with appropriately defined specific diffusion coefficients.  Let  
the coefficient of leaving  ($L$)  the area $ \mathrm{I}$  
toward the area  $ \mathrm{II}$ 
by adepts of the $i$-th population
be $L_{\mathrm{I} \rightarrow   \mathrm{II,}_i}$   and the coefficient of leaving  of the area $ \mathrm{II}$  
toward the area $ \mathrm{I}$ 
by the adepts of 
the $i$-th population be $L_{\mathrm{II}  \rightarrow  \mathrm{I,}_i}$. The corresponding coefficients  for  the
incoming ($I$) followers of the $i$-th population are $I_ {\mathrm{II} \rightarrow  \mathrm{I,}_i}$   
and $I_{ \mathrm{I} \rightarrow  \mathrm{II,}_i}$, or 
$I _{ \mathrm{I} \leftarrow   \mathrm{II,}i}$ and $ I _{ \mathrm{II} \leftarrow   \mathrm{I,}i}$.\footnote{It might be interesting to have
one or several indices attached to $L$ and $I$ in order to distinguish among the migration causes 
like migration currents due to external forces or economic reasons due to gradients in standard of
living, though this would make the notations rather elaborate. Such a remark is highly appropriate
for the cases discussed below. We thank one of the reviewers for such a good suggestion.} 
A necessary balance exists between the number of adepts leaving one area when going into the other, implying 
some obvious relationships between these coefficients. We assume that an adept does not change his/her opinion 
by simply leaving from or going into a region.  
\par
The model equations for the two area ($n+1$) population number evolutions become
\begin{eqnarray}\label{b1}
\rho_{\mathrm{I,}i}^{(t+1)} - \rho_{\mathrm{I,}i}^{(t)} = r_{\mathrm{I,}i}^{(t)} \rho_{\mathrm{I,}i}^{(t)} +
\sum_{j=0}^{n} f_{\mathrm{I,}ij}^{(t)} \rho_{\mathrm{I,}j}^{(t)} + \sum_{j=0}^{n} b_{\mathrm{I,}ij}^{(t)}
\rho_{\mathrm{I,}i}^{(t)} \rho_{\mathrm{I,}j}^{(t)} - \nonumber \\
L^{(t)}_{ \mathrm{I} \rightarrow   \mathrm{II,}i}  \rho_{\mathrm{I,}i}^{(t)} +
I^{(t)}_{ \mathrm{I} \leftarrow   \mathrm{II,}i}  \rho_{\mathrm{II,}i}^{(t)}
\end{eqnarray}
\begin{eqnarray}\label{b2}
\rho_{\mathrm{II,}i}^{(t+1)} - \rho_{\mathrm{II,}i}^{(t)} = r_{\mathrm{II,}i}^{(t)} \rho_{\mathrm{II,}i}^{(t)} +
\sum_{j=0}^{n} f_{\mathrm{II,}ij}^{(t)} \rho_{\mathrm{II,}j}^{(t)} + \sum_{j=0}^{n} b_{\mathrm{II,}ij}^{(t)}
\rho_{\mathrm{II,}i}^{(t)} \rho_{\mathrm{II,}j}^{(t)} - \nonumber \\
L^{(t)}_ {\mathrm{II} \rightarrow   \mathrm{I,}i} \rho^{(t)}_{\mathrm{II,}i} +  I^{(t)}_{ \mathrm{II} \leftarrow   \mathrm{I,}i}  \rho_{\mathrm{I,}i}^{(t)}
\end{eqnarray}
\par
Let us consider, for illustrating the notations,  the particular case of exchange between two areas
when only two populations/opinions are present  in each region: the non-believers $\rho_0$ and
followers of  the (so called prevailing) ideology 1, i.e.  $\rho_1$. The model equations are   reduced to 
\begin{eqnarray}\label{d1}
\rho_{\mathrm{I,}0}^{(t+1)} - \rho_{\mathrm{I,}0}^{(t)} = r_{\mathrm{I,}0}^{(t)} \rho_{\mathrm{I,}0}^{(t)} + 
f_{\mathrm{I,}01}^{(t)} \rho_{\mathrm{I,}1}^{(t)}
+ b_{\mathrm{I,}01}^{(t)} \rho_{\mathrm{I,}0}^{(t)} \rho_{\mathrm{I,}1}^{(t)} - L^{(t)}_{ \mathrm{I} \rightarrow   \mathrm{II,}0}  \rho_{\mathrm{I,}0}^{(t)} +
I^{(t)}_{ \mathrm{I} \leftarrow   \mathrm{II,}0}  \rho^{(t)} _{\mathrm{II,}0} \nonumber \\
\rho_{\mathrm{I,}1}^{(t+1)} - \rho_{\mathrm{I,}1}^{(t)} = r_{\mathrm{I,}1}^{(t)} \rho_{\mathrm{I,}1}^{(t)} + 
f_{\mathrm{I,}10}^{(t)} \rho_{\mathrm{I,}0}^{(t)}
+ b_{\mathrm{I,}10}^{(t)} \rho_{\mathrm{I,}1}^{(t)} \rho_{\mathrm{I,}0}^{(t)} - L^{(t)}_{ \mathrm{I} \rightarrow   \mathrm{II,}1}  \rho_{\mathrm{I,}1}^{(t)} +
I^{(t)}_{ \mathrm{I} \leftarrow   \mathrm{II,}1}  \rho^{(t)} _{\mathrm{II,}1} \nonumber \\
\rho_{\mathrm{II,}0}^{(t+1)} - \rho_{\mathrm{II,}0}^{(t)} = r_{\mathrm{II,}0}^{(t)} \rho_{\mathrm{II,}0}^{(t)} + 
f_{\mathrm{II,}01}^{(t)} \rho_{\mathrm{II,}1}^{(t)}
+ b_{\mathrm{II,}01}^{(t)} \rho_{\mathrm{II,}0}^{(t)} \rho_{\mathrm{II,}1}^{(t)} - 
L^{(t)}_ {\mathrm{II} \rightarrow   \mathrm{I,}0} \rho^{(t)}_{\mathrm{II,}0} +  I^{(t)}_{ \mathrm{II} \leftarrow   \mathrm{I,}0}  \rho_{\mathrm{I,}0}^{(t)} \nonumber \\
\rho_{\mathrm{II,}1}^{(t+1)} - \rho_{\mathrm{II,}1}^{(t)} = r_{\mathrm{II,}1}^{(t)} \rho_{\mathrm{II,}1}^{(t)} + 
f_{\mathrm{II,}10}^{(t)} \rho_{\mathrm{II,}0}^{(t)}
+ b_{\mathrm{II,}10}^{(t)} \rho_{\mathrm{II,}1}^{(t)} \rho_{\mathrm{II,}0}^{(t)} - 
L^{(t)}_ {\mathrm{II} \rightarrow   \mathrm{I,}1} \rho^{(t)}_{\mathrm{II,}1} +  I^{(t)}_{ \mathrm{II} \leftarrow   \mathrm{I,}1}  \rho_{\mathrm{I,}1}^{(t)} \nonumber \\
\end{eqnarray}

From the numerous possible scenarii resulting from the above,  we shall  further 
restrict the  present considerations to    two cases only:
\begin{itemize}
\item
massive leaving of  one region by non-believers  ($i=0$) under the pressure of the followers 
of the prevailing  ideology  ($i=1$);  e.g., protestants leaving Europe because of catholics pressure in the XVI -th century or so. 
\item
massive invasion  of one region by the followers of the prevailing  
ideology  ($i=1$) from  the other region;  e.g.,  catholics conquistadors going to South-America in the XV-th century or so. 
\end{itemize}
 N.B. Although  both cases appear at first rather equivalent, the two cases under interest should nevertheless be suggestively markedly different to 
the  astute reader, because of the birth/death rate conditions imposed on populations $0$ and $1$ respectively , from the mere beginning. 
\subsection{Mass  departure} \label{massleaving}
Let the followers of the prevailing  ideology ($i=1$) put  pressure on the non-believers  ($i=0$) 
in order to fully convert    some region   toward their ideology; thus let  the
a massive leaving of  one region, e.g. $ \mathrm{I}$,  by non-believers  ($i=0$)  toward the other  region, i.e.  $ \mathrm{II}$.
 We assume that no exchange of populations 
between the two regions happens except for the leaving of non-believers from region 
$ \mathrm{I}$ to region $ \mathrm{II}$, i.e. $L^{(t)}_{ \mathrm{I} \rightarrow   \mathrm{II,}0}  \ne 0$ and all other coefficients $L$ and $I$,   
except of course  
$I^{(t)}_{ \mathrm{I} \rightarrow   \mathrm{II,}0} $, in  Eq. (\ref{d1}),   equal to 0. 
\par
 From the first equation of the system  (\ref{d1}), the 
condition for  a decrease in   the number of non-believers ($i=0$),  in $ \mathrm{I}$, is
\begin{equation}\label{d2}
L^{(t)}_{ \mathrm{I} \rightarrow   \mathrm{II,}0}  > r_{\mathrm{I},0}^{(t)} + f_{\mathrm{I},01}^{(t)} \frac{\rho_{\mathrm{I},1}^{(t)}}{\rho_{\mathrm{I},0}^{(t)}} + 
b_{\mathrm{I},01}^{(t)} \rho_{\mathrm{I},1}^{(t)}.
\end{equation}
Eq. (\ref{d2}) tells us that if the coefficient $L^{(t)}_{ \mathrm{I} \rightarrow   \mathrm{II,}0} $ describing the leaving
  is large enough in order to compensate the birth rate and the effect of the
conversion mechanisms, then the population of non-believers in the region $ \mathrm{I}$ will start
to decrease. After some time, the followers of the ideology ($i=1$) can  thus become 
the majority of the population in that region. 
\par
Due to the balance condition, i.e. the conservation of the number of non-believers in both regions, the number of non-believers  
in the region $ \mathrm{II}$ will increase. Thus, from the third equation in  (\ref{d1}), one has similar condition on $L^{(t)}_{ \mathrm{I} \rightarrow   \mathrm{II,}0}$, but relevant for the behavior of populations ($i=0$ and $i=1$) in region $ \mathrm{II}$, i.e.

\begin{equation}\label{d2b}
L^{(t)}_{ \mathrm{I} \rightarrow   \mathrm{II,}0}  \equiv  I^{(t)}_{ \mathrm{II} \leftarrow   \mathrm{I,}0}   > -\frac{ r_{\mathrm{II},0}^{(t)} \rho_{\mathrm{II},0}^{(t)} 
+ f_{\mathrm{II},01}^{(t)}  \rho_{\mathrm{II},1}^{(t)}+ 
b_{\mathrm{II},01}^{(t)} 
\rho_{\mathrm{II},0}^{(t)}
\rho_{\mathrm{II},1}^{(t)} }  {\rho_{\mathrm{I},0}^{(t)}}.
\end{equation}
\subsection{Mass invasion}\label{massinvasion}
Another scenario  leading to some domination of  a region  by an ideology may be due 
to the invasion  of that region from another region by the followers of the 
prevailing  ideology in the latter region. Notice that this scenario is more complex 
than  the scenario of leaving in Sec.  \ref{massleaving} as it requires common efforts  
by the followers of the ideology  in the two considered regions. Let, in Eq. 
(\ref{d1}), $I^{(t)}_{ \mathrm{I} \leftarrow   \mathrm{II,}1}  
\equiv  I^{(t)}_{ \mathrm{II} \rightarrow   \mathrm{I,}1}   \ne 0$ and all other coefficients $L$ and $I$ in (\ref{d1}) 
be $0$,    except of course  $L^{(t)}_{ \mathrm{II} \rightarrow   \mathrm{I,}1} $. The condition for  an increase in the number of believers  ($i=1$),  in region  $ \mathrm{I}$, is
\begin{equation}\label{d3}
I^{(t)}_{ \mathrm{I} \leftarrow   \mathrm{II,}1}  
> - \frac{r_{\mathrm{I},1}^{(t)} \rho^{(t)} _{\mathrm{I,}1}+ f_{\mathrm{I},10}^{(t)} \rho^{(t)} _{\mathrm{I,}0} + 
b_{\mathrm{I},10}^{(t)}  \rho^{(t)} _{\mathrm{I,}1}\rho^{(t)} _{\mathrm{I,}0} 
}{\rho^{(t)} _{\mathrm{II,}1}}.
\end{equation}
In other words, the invasion has to compensate for the death rate (remember that 
$r_{\mathrm{I},1}^{(t)}$) is negative) and  is supported by the conversion mechanisms, through 
$f_{\mathrm{I},10}^{(t)}$ and $b_{\mathrm{I},10}^{(t)}$. Thus, if the followers of an ideology manage 
to organise mass invasion combined with  massive media propaganda (large 
$f_{\mathrm{I},10}^{(t)}$) and large conversation conversion (large $b_{\mathrm{I},10}^{(t)}$) 
and if they  face a "tired" population of  non-believers (small or even
negative $r_{\mathrm{I},0}^{(t)}$) which does not take measures to decrease the number 
of followers of the ideology by conversion (small $f_{\mathrm{I},01}^{(t)}$ and 
$b_{\mathrm{I},01}^{(t)}$) then the invasion could be successful.
The balance condition, i.e. the conservation of the number of ideology ($i=1$) adepts in both regions,  implies that the number of the latter  in  region $ \mathrm{I}$ will increase. Thus, from the fourth equation in  (\ref{d1}), one has similar  condition on
 $L^{(t)}_{ \mathrm{II} \rightarrow   \mathrm{I,}1} $, 
 but relevant for the behavior of populations ($i=0$ and $i=1$) in region $ \mathrm{II}$, i.e.

\begin{equation}\label{d3b}
L^{(t)}_{ \mathrm{II} \rightarrow   \mathrm{I,}1}  > r_{\mathrm{II},1}^{(t)} + f_{\mathrm{II},10}^{(t)} \frac{\rho_{\mathrm{II},0}^{(t)}}{\rho_{\mathrm{II},1}^{(t)}} + 
b_{\mathrm{II},10}^{(t)} \rho_{\mathrm{II},0}^{(t)}.
\end{equation}
\section{Arbitrary number of regions. Further comments }\label{fullmodel}
The final complication of the model with respect to the number of discussed regions is
to consider arbitrary number of regions. We separate the entire studied two-dimensional 
area into $S_k$ regions, $k=\mathrm{I},\mathrm{II},\dots,K$. 
In each region we assume $i=0,1,\dots,n$ populations with densities $\rho_{k,i}$ where $i=0$ denotes
the population of non-believers in any ideology. In general we allow migration of members
of populations between any two of the areas even between those that do not have 
common boundary. The corresponding exchange coefficients are denoted as $L_{k\rightarrow l,i}$ and
$I_{k\leftarrow l,i}$. The system of $K \cdot (n+1)$ model equations  reads\footnote{It is  surely obvious to the reader that a few coefficients are necessarily equal to 0, but all terms are hereby written in order to maintain the number of notations at a reasonable level.}
\begin{eqnarray}\label{c1}
\rho_{k,i}^{(t+1)} - \rho_{k,i}^{(t)} = r_{k,i}^{(t)} \rho_{k,i}^{(t)} +
\sum_{j=0}^{n} f_{k,ij}^{(t)} \rho_{k,j}^{(t)} + \sum_{j=0}^{n} b_{k,ij}^{(t)} \rho_{k,i}^{(t)}
\rho_{k,j}^{(t)} - \nonumber \\
\sum_{l=1}^{K} L_{k\rightarrow l,i}^{(t)} \rho_{l,i}^{(t)} +  \sum_{l=1}^{K} I_{k\leftarrow l,i}^{(t)} \rho_{l,i}^{(t)}
\nonumber \\
i=0,1,\dots,n; \hskip.5cm k=\mathrm{I},\mathrm{II},\dots,K
\end{eqnarray}
This system describes  the dynamics in a $K \cdot (n+1)$ - dimensional phase space where
numerous special points and  complicated attracting manifolds arise, vanish, and move with
  time. The system trajectory can be very complicated as   it travels in a
high-dimensional phase space in order to come close to some  attracting manifold. 
\section{Practical Remarks} \label{remarks}
At this stage, many interesting research questions arise about applications: for  
example,   questions on  the control of the dynamics of the ideological interaction. 
Let us discuss a situation on the control of the spreading of ideologies in a 
selected region of a territory consisting of several sub-regions on the basis of the general model  system 
Eqs.(\ref{c1}). For   bettering such a  control, the dimension of the phase space of the system 
should be reduced. This can happen  by regulating and reducing  the inflow of followers 
of unwanted ideologies,  i.e. technically,  by lowering  the values of $I_{l\leftarrow k,i}^{(t)}$. An 
outflow of followers of unwanted ideologies can be stimulated, if this  does not affect 
the coefficients of mass media conversions.  Otherwise the outflow has to be reduced  also, i.e. 
lowering $L_{k\rightarrow l,i}^{(t)}$. Without significant inflow and outflow in the selected
region, the dimension of the phase space of the ideological dynamics is  practically reduced. 
Within a purpose of control,  one has to deal with the followers of the ideologies 
and with the non-believers in the  region. 
\par
The dynamics of the corresponding populations is determined by the birth and death 
rate of the non-believers as well as by the death rates of the followers of 
all the other ideologies and by the conversion coefficients. Anyone's experience
shows that the conversion by conversation is much part of an e.g. religion conversion in a 
dynamical context \cite{hayw99,hayw05,religion1,religion2,religion566,religion568}, 
while the conversion due to media impact is   well known to be the primary cause of evolution in politics 
\cite{politicalop,politicalopTV,influenceeffectsmassmedia,mediamatters}. Of interest 
is also the conversion of an opinion about goods and consumer products 
\cite{marketing,commercialgoods,wordofmouth,Bass}.  For completeness, recall that external  (propaganda) influence 
and socio-economic pressure can 
be mapped into {\it social forces} \cite{PNAS75.78.4633-7-Montroll-socialforces}. The correspondence with our approach is left for another publication. 
Nevertheless, they would be easily apparent if one was developing the rhs of the complete dynamical equations
\begin{eqnarray}\label{c2}
\rho_{k,i}^{(t+2)} - 2 \rho_{k,i}^{(t+1)} + \rho_{k,i}^{(t)} =  \dots,
\nonumber \\
i=0,1,\dots,n; \hskip.5cm k=\mathrm{I},\mathrm{II},\dots,K
\end{eqnarray}
Thereafter, one can discuss qualitatively cases of recent
ideological control, for   example in   Arab countries and in China. Massive invasion will be considered at an earlier time,  
i.e. the XIV-th century invasion of Europe by Turks.
\par
Recently,  the failure of   ideological control has been seen in several Arab countries.
Several reasons for this come in mind on the basis of the discussed model. Let in this case the index $i$ in the model equations
denotes: $i=0$ - no ideology; $i=1$ - official ideology; $i=2$ - Islamism; $i=3$ - liberal ideology; $i=4$ - moderate Islam. 
The official ideology connected to  
corresponding dictatorship regimes has gained followers mostly through the domestic mass media
propaganda and conversations supported by financial support.\footnote{There is also the ''service and welfare support'' to the need of the population, - 
which is not really mass media propaganda nor conversation, but rather an opinion ''formation'' due to some "theoretically gratuitous  
altruism''  \cite{[r71],[rPB73.70],[rPB91.82],[r11.EHB24.03],[r13.N425.03]} by ''charitable
organisations'' following resources transfer, with political goals \cite{Kuranjusticeislam,Ayubipoliticalislam,GulalpwelfareTK,Bayat,egyptislampolitics}.
In view of its clustering effect, somewhat a group response to group action, it might be considered to be   a higher order term in 
Eqs.(\ref{a2}), or still be integrated into each   $f_{ij}^{(t)} $   coefficient.} 
This leaded to significant positive value of $f_{10}$. $b_{10}$ had significant positive value too. 
Islamists were not  successfully converted to the official ideology, i.e., the values of $f_{12}$ and $b_{12}$
were small and negative. More successful was the conversion from the liberal to official ideology which means that $f_{13}$ and $b_{13}$
had positive values. Relatively successful was the conversion from Islam to
the official ideology (as the official ideology was partially based on Islam). As consequence of this the coefficients
$f_{14}$ and $b_{14}$ had significant positive values.  Thus the strategy of the corresponding government was to
convert as many as possible non-believers and followers of the Islam to the official ideology and to decrease as much as possible
the conversion of the followers of the official ideology to Islamism.
As well as the government had enough money and propaganda power it was possible to follow such a strategy.
\par
The liberal ideology spread by the mass media was mainly connected to the west technology: Interned and West TV
channels  and military shows. This ideology was not able to compete successfully against the government ideology which means that the
corresponding conversion coefficients  $f_{30},b_{30}$ had smaller  values in comparison to $f_{10},b_{10}$.
The conversion from official ideology to liberal ideology was not  successful, i.e., $f_{31}$ and $b_{31}$ had negative values.
The values of $f_{32}$ and $b_{32}$ were very small negative (there was  no conversion of Islamists to liberal ideology and small number
of followers of liberal ideology converted to Islamism). The conversion of the followers of the Islam to liberal ideology was also
not successful. Thus the values of $f_{34}$ and $b_{34}$ were almost zero. Thus the strategy of the followers of the liberal ideology was
to convert the non-believers (mainly young people) and to minimize the decreasing of the followers by conversion to the official ideology and
Islamism.  
\par
The Islamism gained  followers by conversations, hypocritical altruism and the domestic mass media.
Using the power of the Islam large values of $f_{20}, f_{24}$ and $b_{20},b_{24}$ have been maintained. $f_{21}, b_{21}$ had small positive values
as some supporters of the official ideology converted to Islamism. The values of $f_{23}$ and $b_{23}$ were also positive but much smaller
as very few supporters of the liberal ideology converted to Islamism.
\par
The followers of the moderate Islam  increased in the course of years because of the large conversion rates $f_{40},b_{40}$
from (youth) non-believers to followers of the Islam. Much smaller conversion happened from Islam to followers of the official ideology (small
negative values of $f_{41}$ and $b_{41}$) and even smaller was the conversion from the followers of the liberal ideology
(very small positive values of $f_{43}$ and $b_{43}$). The conversion of fanatic Islamists to the moderate Islam was not successful (negative
values of $f_{42}$ and $b_{42}$). The centuries old strategy of the followers of moderate Islam to convert very successfully (young) non-believers
worked very well and compensated the losses due to conversion to official ideology and Islamism.  
\par 
The outflow  of people from these countries  toward others was not (at this time)  large enough, though the large birth rates in the Arab countries produced 
a large amount of jobless youth which becomes resistive to accept the official ideology in the corresponding Arab country. Conversion to the corresponding official  
ideology by conversation,  was apparently not very successful  either and the values of $f_{10}$ and $b_{10}$ remained positive but decreased very much with
respect to the values of the corresponding coefficients for the other ideologies. Thus in the course of the years
the non-believers  were converted much more successfully to moderate Islam, Islamism and even to the liberal ideology.  The spreading of  liberal opinions  
following mass media conversion, such as Internet, Facebook, Twitter, etc., was not 
totally controlled and thus the non-believers were even more resistive for conversion 
to the official ideology. Thus the official (and mostly totalitarian) ideology in 
these countries had decreased percentage of followers and finally collapsed. 
Now in these countries we observe fast growth of the fanatic Islamists, many moderate followers of Islam,  still influential
population of followers of the former official ideology and  a slowly increasing amount of people 
that ''convert'' to  some liberal ideology, under the influence of the mass media from the West. 
Then the future development in these societies in the next years will depend much on the competition 
between the liberal ideology, Islamists and moderate Islam followers. The liberal ideology has a problem: 
it can not convert very successfully the
followers of the moderate Islam. The Islamists do not have this problem.
 Thus a possible capsulation of the Arab societies (and especially the 
decreasing influence of the West mass media) could lead to  a
dominance of Islamists and Islamic ideology which is quite deeply rooted in these societies and has  a large 
tradition  of conversion of non-believers.
\par
Much more successful is the ideological control in China.  Many elements of the 
capitalist economic system help for successfully absorbing  the inflow of young 
people  who have studied in the West. This absorption is easily transformed  by  
(social or economic or political) pressure conversion in   the official Chinese 
ideological system. Birth rates are much under control and the outflows of followers 
of different ideologies are much regulated \cite{TienPopBul83}. 
\par
Let in the case of China the value of the index $i$ denote ideologies as follows: $i=0$ - no ideology; $i=1$ -
official ideology; $i=2$ - liberal ideology.
Mass media is under control and the Chinese government takes steps toward controlling 
the Internet and the connections between  social networks. This ensures large positive
values of the conversion coefficients $f_{10}$ and $b_{10}$ whereas the values of the
corresponding coefficients for the liberal ideology $f_{20}$ and $b_{20}$ are positive but
much smaller. As the official ideology contains element of liberalism the values of
conversion coefficients $f_{12}$ and $b_{12}$ are small positive and the values of
corresponding coefficients for the liberal ideology $f_{21}$ and $b_{21}$ are negative.
We note that in principle the conversation conversion is much out of control\footnote{Some 
investigation in such a respect is worth to be 
pointed out: in the Moss model \cite{moss1}, social   embededness takes the form of
observation of neighbours public consumption activities such as
garden watering and car washing as well as word of mouth
communication.}
but up to now it is successfully countered by  governmental propaganda and  internal 
social pressure and the value of $b_{20}$ is small. Thus the ideological
situation in China is stable and the ideological dynamics is dominated by the 
official ideology. As no large demographic problem  is  expected in the next decades 
and if the economic development continues, the stability of the ideological situation 
is China is expected to remain largely unchanged.   
\par
Finally, from the historic situations connection to mass invasion of followers of some
ideology to a region dominated by another ideology let us consider the invasion of
the Ottoman Turks into Europe that started from the second half of the 14-th century.
We consider two regions: region I.: the Balkans (the invaded region); and region II.
(Anatolia, Asia Minor and surrounding areas of Asia). 
At this time the invaded region: the Balkans were populated by followers of
Christianity (mostly from the orthodox branch but with presence of followers of 
catholicism too). Weakened by the 'Black death' plague the small Balkan countries
were conquered one after another by  the Muslim Ottoman turks who started to
impose  Islam, ruining churches or transforming these into mosques.
Let us discuss qualitatively the situation from the point of view of the model presented above.
Let the values of the index $i$ mean the following: $i=0$ - non-believers; $i=1$ -
Christians; $i=2$ - followers of the Islam. The birth rates supported the
population of non-believers that in course of years converted to Christianity or Islam.
At that time there was no mass media as we know it today  so $f_{{\rm I},01}$, $f_{{\rm II},01}$, $f_{{\rm I},10}$, 
$f_{{\rm II},10}$ were practically 0 and there was no
possibility of media influence on non-believers in order to convert them to Christianity.
The values of the coefficients $f_{{\rm I}, 02}$, $f_{{\rm II}, 02}$, $f_{{\rm I}, 12}$, 
$f_{{\rm II}, 12}$ were practically 0 too as there was no conversion of the followers of Islam to Christianity
or to the population of non-believers. The public
spreading of the administrative orders of the Ottoman administration and the lower taxes
for Muslims can be treated as sort of media that contributed to conversion and led to small positive values
of the coefficients $f_{{\rm I},20}$, $f_{{\rm II},20}$,$f_{{\rm I},21}$  and $f_{{\rm II},21}$. 
The main process for ensuring the existence of
the Christian population was the conversion of (young) non-believers to Christianity by
means of conversation, i.e., the value of $b_{{\rm I},10}$ was large positive. The
value of $b_{{\rm II},10}$  was small positive as in the region II the Christian population was
surrounded by large Muslim population that put pressure acting in the direction of muslimisation
of the non-believers in that region. This was a branch of the Ottoman assimilation mechanism that
included displacement from Europe to Asia and then muslimisation. In the two regions there was
almost no conversation conversion from Christianity or Islam to the population of non-believers. Because of this
the values of the coefficients  $b_{{\rm I},01}$, $b_{{\rm I},02}$, $b_{{\rm II},01}$, $b_{{\rm II},02}$ 
were negligibly small. The conversation conversion from Islam to Christianity was also negligibly small, i.e.,
the values of the coefficients $b_{{\rm I},12}$, $b_{{\rm II},12}$  were practically $0$.
\par
The politically dominating Muslim population was not able to keep large enough level of conversation
conversion in the region ${\rm I}$: the values of $b_{{\rm I},20}$ and  $b_{{\rm I},21}$ were positive but far away
from the needed values for fast conversion of the Christian population to Islam. The population of
entire regions refused to convert to Islam. 
For increasing the value of $f_{{\rm I},20}$ the '$enichars$' tactic was used for almost three centuries - Christian children were
taken away from their parents and then converted to Islam. The increase of the values of $f_{{\rm I},21}$ and $b_{{\rm I},21}$
was attempted by campaigns to muslimise entire regions of the Balkan peninsula (todays Albania where
much more than the half of population were catholics, todays Bosnia and Herzegovina, and todays Rodopa mountain
region in Bulgaria where the population consisted of almost 100 \% followers of Orthodox Christianity). 
Who refused to convert was killed. In the region II (Asia) the situation was different as there the Muslim
population was large majority and the pressure for conversion of Christian population was effective, i.e., the
values  of  $b_{{\rm II},20}$ and  $b_{{\rm II},21}$ were positive and large enough for the conversion to happen. 
\par
Important point of the Ottoman strategy for muslimisation of the Balkans was connected to migration.
There was massive migration of Muslims from region II (Asia) to region I (Balkans) especially after
Timur (known also as Tamerlane)  defeated and captured the Ottoman sultan Bayezit I in 1402 near todays Ankara. The
remains of the Ottoman army were transported from Anatolia (region II) to Thrace (region I) by Venetian
ships.\footnote{Venetians preferred to have the weak (at that time)  Ottoman turks as enemy and a buffer against  the mighty Timur but 
Timur aimed to restore the Mongolian Empire of Chengis Khan and turned to conquer China. The  Ottoman army could reorganize and crushed
the Bulgarian uprising in 1405. Thus   the way to Europe was opened for the Ottoman
that managed to besiege even Vienna two times in the next 300 years.} Further
migration followed and the coefficients $L_{{\rm II \to I},2}$ and $I_{{\rm I \leftarrow II},2}$  had large positive values. 
The migration of Muslims in opposite direction was very small and the values of the coefficients $L_{{\rm I \rightarrow II},2}$ 
and $I_{{\rm II \leftarrow I},2}$ were practically $0$.

\par
Parallel to the migration of the Muslim population  the Ottoman turks displaced the Christian population and non-believers (children) 
and part of it was sent from region I to region II (example of this is the displacing of large part of Christian population of Belgrade to Istanbul) so the
coefficients  $L_{{\rm I \to II},1}$,  $L_{{\rm I \to II},0}$,  $I_{{\rm II \leftarrow I},0}$,  $I_{{\rm II \leftarrow I},1}$  had also large positive value. 
Such population usually remained in Anatolia and had been converted to Islam through the centuries of Ottoman rule. Thus the values of the 
coefficients  $L_{{\rm I \to I},1}$,  $L_{{\rm II \to I},0}$,  $I_{{\rm I \leftarrow II},0}$ and  $I_{{\rm I \leftarrow II},1}$  were very  small.  
\par
The conversion strategy of the Ottoman turks acted for a very long time.
Because of this,  450 years after the start of the invasion between 30\% and 50\% of the population
of several regions of the Balkans has been  practically converted to  Islam.
\section{Conclusions} \label{conclusions}
In this paper we have discussed a discrete model of ideological interaction which 
is easy for practical implementation for prognosis of the dynamical evolution of ideological 
situations.  The model system is based on a discrete time process within a Lotka-Volterra framework, but 
allowing for migration of populations between areas. The model contains specific terms  mimicking as much 
as possible stylised facts or, as usually, common sense in opinion evolution dynamics.  Each country or 
region of countries can be separated into a number of 
regions for which some data on the number of followers of different ideologies can be 
collected. Birth and death rates are often available from
statistical sources indeed. From these sources estimations of the coefficients of  mass 
leaving and incoming can be usually 
obtained  as well.  What remains, and is  the most difficult task,  is the determination 
of the conversion coefficients. However  it seems that they can be  obtained by standard 
or appropriate
statistical procedures,  like opinion pool evaluations, or by indirect means of behaviour observation. Of course long time interval 
observations are required in order to determine the time change of the
coefficients  for model equations. 
\par
After determining the coefficients one can launch the model, starting from Eqs.(\ref{c1}), with known or appropriate available initial conditions. Under the
assumption that the coefficients of the model equations vary  rather slowly, it  can  be fairly assumed that  either they are constant 
during several time steps or   they change following a linear law  during these  steps. 
Therefore, the model is expected to give a valuable approximate picture of the ideological situation in the observed region
for these time steps. Of course, such approximations can be best applied when the situation in the corresponding
regions is stable and  when no large social change  happens. In  times of large social changes, and extreme events, like wars, revolutions,  ...,  the
constant or   linear approximation   can be  unsatisfactory. 
\par
As  discussed here above, the model admits  a very large number of possible scenarios.  
In conclusion, even without solving  in detail the system of equations, one can obtain some orientation about  ideology situations
in different regions  merely  on the basis of qualitative discussions of the equations of the model. Only limited to 
opinions, it can be fruitfully compared with attempts of  the so called  World Model  \cite{Meadows} for socio-economic considerations.

\section*{Acknowledgments}
We would like to acknowledge the stimulating discussions with many colleagues
participating in the COST MP0801 Action 'Physics of competition and conflict'. 
We thank the COST MP0801 Action for the support of our research.  We thank
the reviewers for questions allowing to improve the manuscript.

\end{document}